\documentclass{ptapap}
\usepackage{color}
\usepackage[title,titletoc,toc]{appendix}
\usepackage{float,lscape}
\newenvironment{myquote}{\setlength{\leftmargin}{1em}\color{blue}\quotation}{\endquotation}
\definecolor{todo}{rgb}{0.8,0.2,0.3}
\author{Andrzej Pigulski}[IAUWr]
\author{the BRITE Team}
\affil[IAUWr]{Instytut Astronomiczny, Uniwersytet Wroc{\l}awski, Kopernika 11, 51-622 Wroc{\l}aw, Poland, e-mail: pigulski@astro.uni.wroc.pl}
\title{BRITE Cookbook 2.0}

\begin{document}
\maketitle
\begin{abstract}
The raw BRITE photometry is affected by the presence of many outliers and instrumental effects. We present and discuss possible ways to correct the photometry for instrumental effects. Special attention is paid to the procedure of decorrelation which enables removal of most of the instrumental effects and considerably improves the quality of the final photometry. 
\end{abstract}
\section{Introduction}
After release of the first data obtained by the BRITE nano-satellites (hereafter briefly called `BRITEs'), it became clear that the users of BRITE data will need a document introducing a novice to the mission and its data. In this way, the idea occurred\footnote{The idea was originally proposed to the author by Dr.\,Gerald Handler.}  to write a `BRITE Cookbook', which summarizes the up-to-date experience in working with the real BRITE data and shows step by step how to get a scientifically useful time series. The first version of the Cookbook was completed in May 2015 and made available via the BRITE Photometry Wiki page\footnote{http://brite.craq-astro.ca/doku.php?id=cookbook. The last previous version, 1.6, refers to Data Release 2; hereafter DR2. The present version of the Cookbook is suitable for all data releases issued by now (DR2, DR3, DR4, and DR5) and hopefully also the two planned ones (DR6 and DR7).}, one of the primary sources of the information on BRITE data, reductions, and targets. The growing experience in working with BRITE data, the much better understanding of instrumental effects, and the changes of BRITE data formats made the last version of the Cookbook outdated. This is the main motivation for the present paper, which includes an updated Cookbook. 

In the meantime, the mission itself has been described in detail in a series of technical papers. In addition to the introductory paper on the BRITE mission \citep[][hereafter Paper I]{2014PASP..126..573W}, two papers explaining technical details of the mission were published \citep[][hereafter Paper II and III, respectively]{2016PASP..128l5001P,2017A&A...605A..26P}. All three papers are very important sources of information on the BRITE mission, the data, and the final photometry. Therefore, we recommended to read them before starting to work with BRITE data.

Other useful sources of information on BRITE and the data are the following:
\begin{itemize}
\item BRITE-Constellation page (http://www.univie.ac.at/brite-constellation/)
\item BRITE Photometry Wiki page (http://brite.craq-astro.ca/doku.php?id=start)
\item BRITE Public Data Archive (https://brite.camk.edu.pl/pub/index.html)
\end{itemize}

The present Cookbook reflects the experience not only of the author but also of many people that were involved in the work with BRITE data, especially the members of the PHOtometry Tiger Team (PHOTT) and the Quality Control Team (QCT), both led by Dr.~Herbert Pablo. All members of these two teams and all BRITE data users are greatly acknowledged for contributing their experience in working with BRITE data. 

The main purpose of this paper is to show how to proceed with the raw BRITE data to get scientifically useful time series. The problem is that the photometry, which is produced by the reduction pipelines presented in Paper III, is affected by several instrumental effects. Therefore, some actions need to be undertaken to get rid of them. However, {\sl there is no single best way to work with any data} and BRITE data is not an exception. The comments given below should be regarded only as hints, not as the only possible way to proceed. In general, working with BRITE data requires some level of interaction and flexibility. Even if authors have a script which can be run to produce the final result within a few minutes, they typically run it several times to inspect some intermediate plots and decide on the optimal parameters. The proposed procedure is therefore rather iterative and interactive.

\section{A few words about BRITE-Constellation}\label{fwords}
Before we start a description of the analysis of BRITE photometry, we present some very general characteristics of the mission, data and photometry. More details can be found in Papers I\,--\,III.
\begin{itemize}
\item {\sl Satellites.} The set of BRITE nanosatellites is called BRITE-Constellation and consists of five working\footnote{The sixth BRITE, Canadian BRITE-Montr\'eal (BMb), did not separate from the upper stage of the launcher for unknown reason.} low-orbit satellites launched in 2013 and 2014. Two host blue filters, the other three, red filters. Some characteristics of the BRITEs (names, abbreviations, launch dates, and orbital periods) are given in Table 6 of Paper II.
\item{\sl Field of view.} The field of view of the BRITEs is slightly affected by vignetting and covers approximately 24$^\circ$\,$\times$\,20$^\circ$ in the sky\footnote{BHr has different optics (four instead of five lenses) and a slightly smaller field of view.}; see Fig.\,1 in Paper III.
\item{\sl Images.} Full-frame images are downloaded very rarely, in particular, at the beginning of the commissioning phase.\footnote{Recently, full-frame images were downloaded for each BRITE to investigate CTI-related instrumental effects (the CTI is explained in Sect.\,\ref{brphot}).} Typically, only small parts of the image called subrasters, which include pre-selected stars, are downloaded. This procedure is dictated by the limits of the amount of data that can be transmitted to the ground stations. 
\item{\sl Subrasters.} The images are intentionally defocused to avoid saturation and to decrease the dependence of photometry on pixel-to-pixel sensitivity variations. The rasters are either square (stare mode of observing\footnote{See Papers II and III for a detailed explanation of the modes of observing.}) or rectangular (chopping mode of observing). A typical size of a subraster is equal to 28\,$\times$\,28 (stare mode) or 48\,$\times$\,28 pixels (chopping mode). On average, between 20 and 30 stars are observed in a single field.
\item{\sl Observing strategy.} Selected fields (mostly along the Galactic plane) are observed typically by two or three BRITE satellites to secure two-colour observations. If only relatively faint targets are available in a selected field, a single red-filter satellite run is scheduled.
\item{\sl Modes of observing.} All satellites began observing in the stare mode, in which satellites did observations in a fixed position. When a problem with the increase of the number of hot pixels was recognized, a new mode of observing was proposed by Dr.~Adam Popowicz. It is based on a combination of nodding between two slightly offset pointing positions and subtraction of consecutive images. Since the new mode resulted in a better photometry, all BRITEs switched to this (chopping) mode of observing since mid-2015. In the chopping mode, subrasters are larger than in the stare mode and rectangular.
\item{\sl Exposure times.} The exposure times for BRITE images range between a fraction of a second and 7.5~s. The most typical exposure time is 1~s. Consecutive exposures are separated by about 20~s. Typically, the observations cover 10\,--\,20~minutes of each $\sim$100-min satellite orbit.
\item{\sl Photometry.} The current BRITE photometry is aperture photometry with either constant or thresholded (non-circular) aperture. A detailed description of the photometric pipelines is given in Paper III.
\item{\sl Data releases.} At the time of writing (December 2017), data are available as four different Data Releases (DR2 to DR5). DR2 comprises the photometry of the stare-mode observations and a part of the chopping-mode observations in the Per~I field. The remaining three data releases, DR3\,--\,DR5, resulted solely from the chopping-mode data. The releases differ in the number of parameters provided; see Appendix A in Paper III for a detailed description.
\end{itemize}

\section{BRITE photometry}\label{brphot}
The two photometric pipelines used to obtain BRITE photometry are described in Paper III, so that only a few general comments are provided here. Already at the beginning of our work with BRITE images, it was obvious that their reduction would not be a trivial task. The main factors that pose problems in the reduction are the following:
\begin{itemize}
\item The number of hot (and associated but weaker cold) pixels and other chip defects (bad columns) is large and growing with time. This is mostly a result of bombardment of the detectors by cosmic-ray protons. In the lack of effective shielding, they cause chip defects.
\item Although satisfying mission requirements, the tracking is sometimes not ideal, which results in a permanent wobbling of a star in a raster.  In some images, stars even fall partially beyond the raster, making such images inappropriate for aperture photometry. The drift of stars may cause considerable smearing even in 1-s exposures. This is the most severe factor diminishing the quality of BRITE photometry.
\item In some regions of the BRITE detectors, an additional defect called charge transfer inefficiency (CTI) occurs. CTI causes additional smearing. CTI regions are mainly caused by low-incidence-angle cosmic-ray protons, while hot pixels are mainly the result of more perpendicular incidence, depending also on the proton energy.
\end{itemize}

All three factors degrade the final photometry. A lot of effort was aimed at minimizing their influence. First of all, when the problem was first recognized, four BRITEs were not yet launched. It was therefore decided to shield the detectors in three of them (BMb, BTr and BHr). Observations indicate that the shielding is most effective in BHr, where the different optical design allowed for more space to place the boron shield behind the detector. Next, the satellites switched to the chopping mode of observing. The first BRITE data based on chopping mode of observing were delivered in DR2 for the Perseus field. By now, BRITE data for 452 stars\footnote{An updated list of stars observed by the BRITEs and the status of the data can be found in the BRITE Wiki page.} in 23 observed fields (some re-observed) were released.

The raw BRITE photometry is affected by several instrumental effects which result in a large number of outliers and larger-than-expected scatter of the data. These effects are described in a separate article entitled `Instrumental effects in BRITE photometry' (Pigulski et al., these proceedings, hereafter Paper IV). The present Cookbook focuses on ways to remove these effects, most importantly, on decorrelations and outlier removal, which are the most important procedures that need to be performed to get data suitable for time-series analysis.

\section{Data files}\label{bdata}
\begin{figure}[!ht]
\centering
\includegraphics[width=0.74\textwidth]{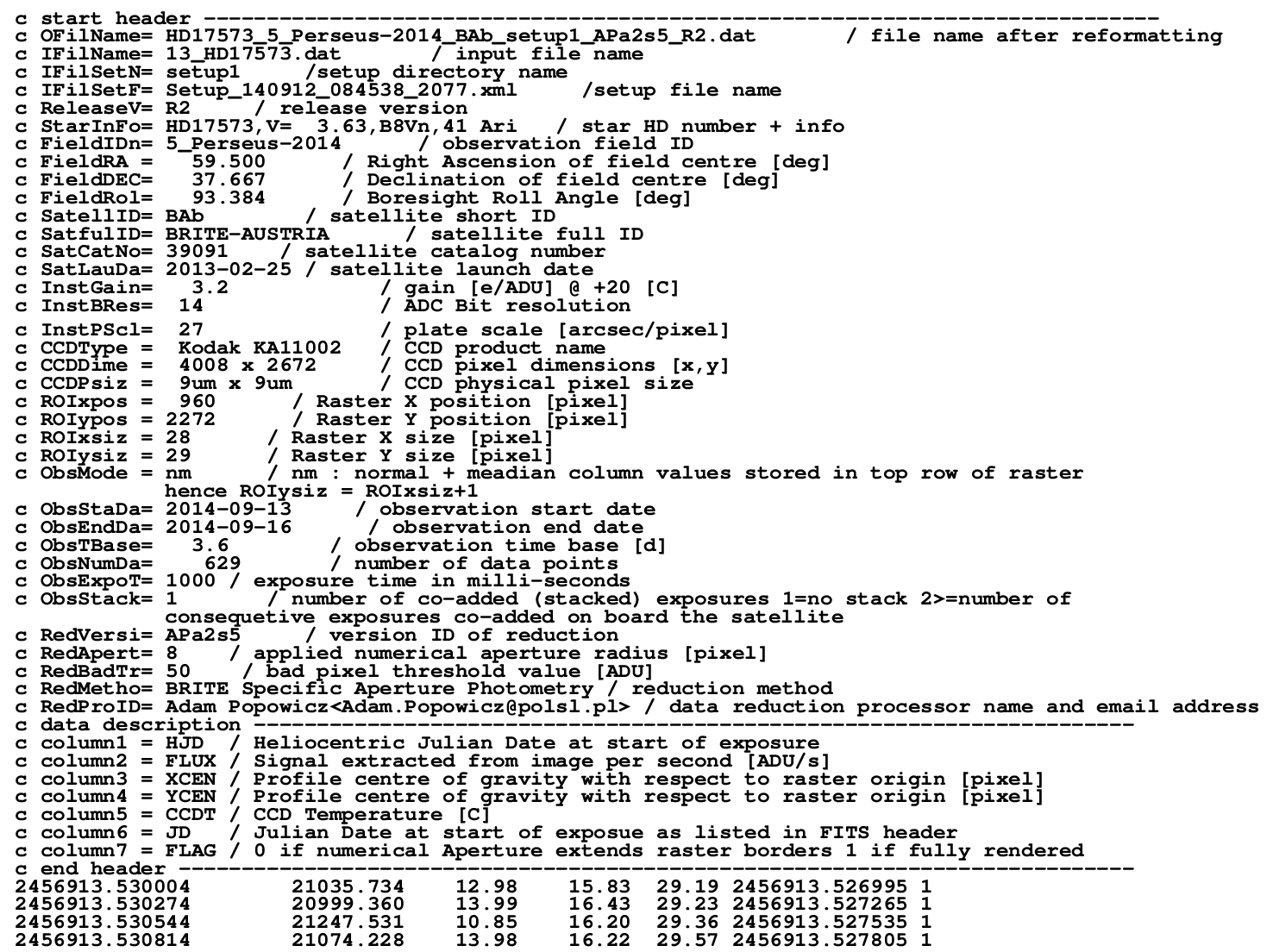}
\caption{The header and the beginning of the data section for a sample DR2 file.}
\label{data}
\end{figure}

The BRITE data are sent to users as ASCII files separately for each BRITE satellite, observed field and observational setup\footnote{The observational setups can differ for various reasons, the most frequent being the change of the raster size or its position in the CCD; see Paper III for the explanation of setups.  Because each difference may lead to some non-identical systematic effects, data are reduced separately for each setup. Eventually, they can or even should be combined, but one may want to first correct them separately for the instrumental effects.}. The files contain a header section followed by a data section in the form of 7 to 12 columns, depending on the DR; see Appendix A in Paper III for the full description. The header records start with the letter `c' and include all important information on the satellite, observed field, and observational setup. The data segment contains numbers in columns; the column entries are briefly explained in the bottom part of the header.  A sample beginning of a DR2 data file is shown in Fig.\,\ref{data}.

In all DRs, the sixth column contains the Julian Day (JD). This seems to be redundant since the Heliocentric Julian Day (HJD) is given in the first column. However, JD can be used to identify the corresponding image since JDs are used as file names of the raw images. Presently, the raw images are not publicly available, but this is planned for the future, so that this information is preserved. {\sl The data sets do not contain information on the uncertainty of the measured flux.} Since some fitting programs require uncertainties, we will show later how to derive and add such values to the data files. 

\section{Preparatory steps and removal of outliers}
The raw BRITE light curve is a combination of intrinsic variability (if present) and instrumental effects. As pointed out above, the main purpose of the Cookbook is to help the user to remove (or at least minimize) the latter in order to obtain time-series photometry free of instrumental effects. This is needed to use this photometry for the scientific purposes. Therefore, we will focus on the ways to separate (and remove) the instrumental effects from the intrinsic variability.

\subsection{Reformatting}
First, users need to decide whether they want to work with fluxes or magnitudes. This is a fully subjective choice and --- in principle --- both choices are equally good. Personally, we prefer to work with magnitudes, so we start with converting fluxes to magnitudes. In addition, the working file should not have a header, so that we remove it. The users may also want to remove some columns containing auxiliary data or flags. There is an infinite number of ways to do this using e.g.~the shell commands (the `{\tt grep}' and `{\tt awk}' commands are very useful in this context). For example, the part of our {\tt tcsh} script which does reformatting for a DR2 file looks like this (the starting `$>$' is a terminal prompt, `lc.data' is a working file name, `lc.tmp' is a temporary file name):
\begin{myquote}
{\footnotesize\tt
\noindent
$>$ cp [original\_file\_name] lc.data

\noindent
$>$ grep -e ObsExpoT lc.data | awk '\{printf("\%12.6lf{\textbackslash}n",\$3/(2.0*1000.0*86400))\}' 

> tmp.texp

\noindent
$>$ set TEXP=`awk '\{print \$1\}' tmp.texp`

\noindent
$>$ grep -v c lc.data $>$ lc.tmp; mv lc.tmp lc.data

\noindent
$>$ awk -v mv=\$\{TEXP\} '\{printf("\%12.6lf \%12.6lf \%s \%s \%s{\textbackslash}n",\$1-2456000.0+mv,

-2.5*log(\$2)/log(10.0)+14.5,\$3,\$4,\$5)\}' lc.data > lc.tmp 

\noindent
$>$ mv lc.tmp lc.data
}
\end{myquote}
The {\tt awk} command shows that we also subtract 2456000.0 from the HJD, but add half of the exposure time (the original HJD is given for the beginning of the exposure). We also add an arbitrary constant of 14.5 to the derived magnitudes, i.e.~$\mbox{magnitude} = -\mbox{2.5}\log\mbox{(FLUX)} + \mbox{14.5}$. After this procedure, a single record in the working file {\tt lc.data} looks as follows (the columns with JD and FLAG were not copied):
\begin{myquote}
{\footnotesize\tt
820.812838  \quad  2.092845  \quad  14.13  \quad  13.29 \quad15.83}
\end{myquote}

\begin{figure}[!ht]
\centering
\includegraphics[width=0.75\textwidth]{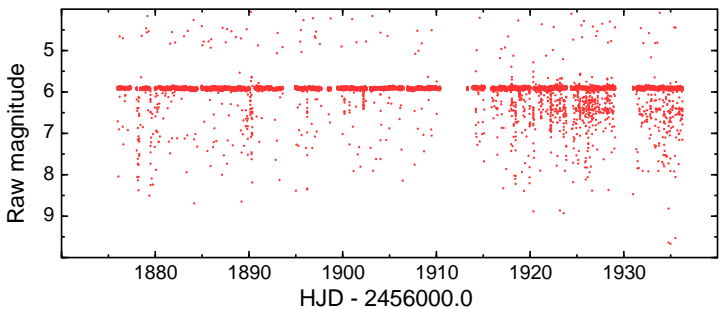}
\caption{\small Raw DR5 BHr light curve (setup 8) of HD\,92287 (V514~Car) in the Car~I field.}
\label{raw-lc}
\end{figure}

This is something one may want to plot. Again, one can use a favourite plotting tool (our choice is {\tt Gnuplot}\footnote{http://gnuplot.sourceforge.net}) to do this. The result may look like the light curve shown in Fig.\,\ref{raw-lc}. At first glance it looks horrible (range in magnitude is about 10~mag), but don't panic! A relatively large number of outliers is typical for BRITE data and we will remove them soon.

\subsection{Cutting off extreme outliers}\label{cuts}
Figure \ref{raw-lc} shows that the raw BRITE light curves are strongly affected by the presence of a large number of outliers. They originate from the subrasters in which the stellar image is located close to its edge (or even absent if a satellite lost fine pointing), the image is extremely smeared by the satellite movement, or the subraster is affected by CTI. Looking at Fig.\,\ref{raw-lc} it seems obvious to cut off the outliers in the next step. In fact, it is good to remove not only the outliers in magnitude, but look at the distribution of all parameters and cut off, step by step, all data with extreme values of the parameters. This is important from the point of view of the subsequent decorrelations and will be discussed in Sect.~\ref{decor}. We would like to point out here, however, that in order to define a reliable correlation function, it is better to have the whole range of a given parameter densely populated with data points. This argument justifies cutting off the extreme values. Usually, we reject only a small fraction of the data, which are of lower quality because reliable decorrelation cannot be done in a scarcely populated parameter range.
\begin{figure}[!ht]
\centering
\includegraphics[width=0.98\textwidth]{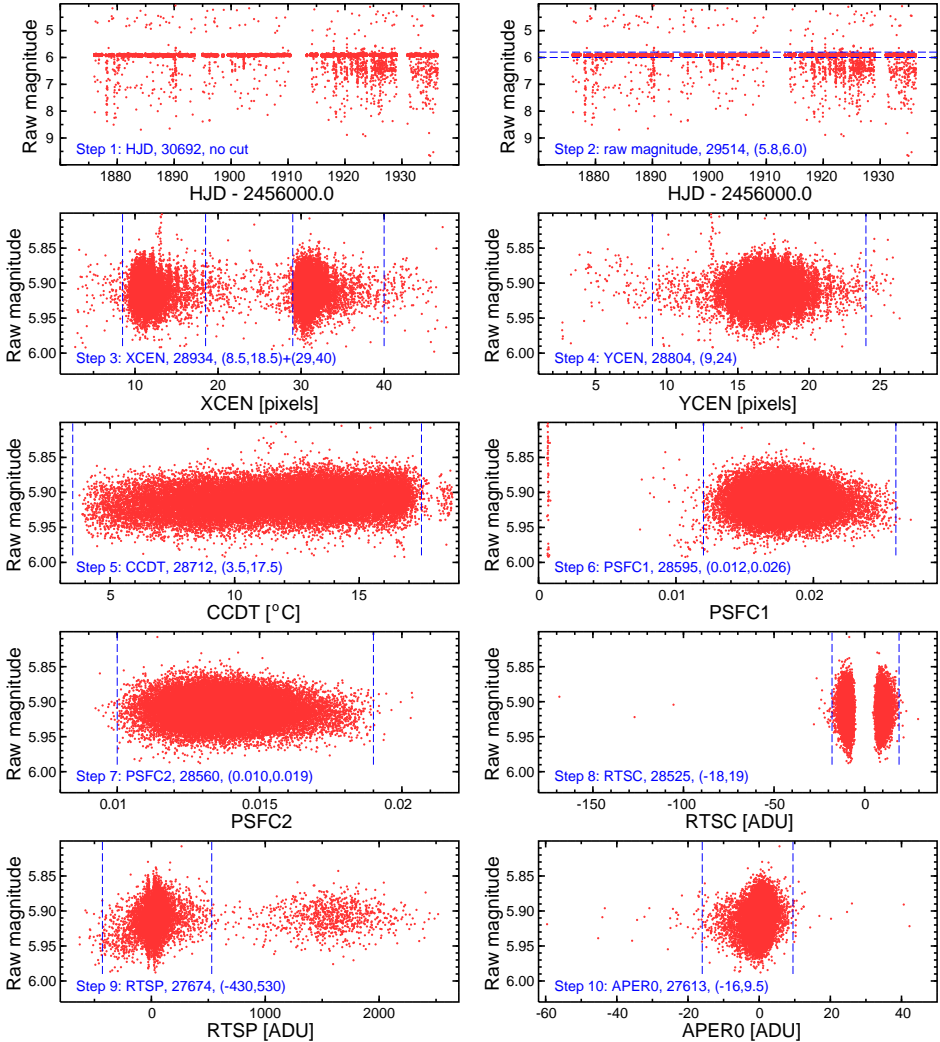}
\caption{A sequence of the procedure of filtering out the data with extreme values of time, magnitude and the eight decorrelation parameters as labeled along each abscissa and explained in Paper III. As an example, the same data set as shown in Fig.\,\ref{raw-lc} was taken. For each step, the number of points left after cutting and the cutting range(s) are shown. The dashed lines show the adopted cutting limits.}
\label{cut-off}
\end{figure}

The procedure of eliminating the extreme values of all decorrelation parameters, magnitude and time is shown step by step in Fig.\,\ref{cut-off} for the same data set as in Fig.\,\ref{raw-lc}. In fact, there is no need to truncate the data in time (HJD, step 1)\footnote{One may, however, do this for other setups and cut off the part of the data, which is much worse than the rest, e.g.~data affected strongly by CTI.}. One may also consider leaving the cloud of data points for RTSP $\in$ (1000, 2200), step 9, which are deleted in the present example. On the other hand, the need for some other selections is obvious, especially that in step 2 (in magnitude)\footnote{Cutting off the outliers in magnitude should be performed only after a visual inspection of the light curve in order not to remove genuine short brightenings (outbursts) or dimmings (eclipses). It is always better to retain some outliers than to reject the most valuable features in the light curve.}, PSFC1 (step 6), or RTSC (step 8). For the presented example, the whole procedure keeps 27613 out of the original 30692 data points, i.e.~the filtering removes about 10\% of the data. 

Step 3 of the cutting procedure (for XCEN, the $x$ coordinate of the stellar centroid) requires a comment. The two clouds of data points correspond to the two chopping positions of the star in the subraster. The area in between is scarcely populated by the data points because the star is placed on either one or the other side of the subraster. Thus, it makes sense to cut off also the data points which fall in between. 

The photometry in the chopping mode is made on a star in two different positions in a CCD and this may result in a magnitude offset between the two positions. One may therefore consider splitting the data into two parts according to the position in chopping, perform the subsequent decorrelations separately for each position and then merge the data back. A detailed investigation if this procedure leads to a noticeable improvement of the photometry, has not been made, however. 

\subsection{Removing the remaining outliers}\label{rem-out}
\begin{figure}[!ht]
\centering
\includegraphics[width=0.65\textwidth]{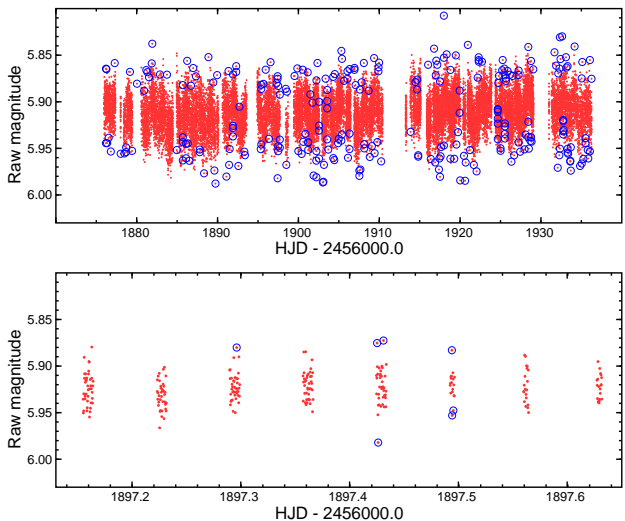}
\caption{Top: The BHr light curve of HD\,92287, the same data as shown in Fig.\,\ref{cut-off}, after removal of extreme outliers. Bottom: A part of the data shown in the upper panel. In both panels, the data marked as outliers by the GESD algorithm (see text) are encircled.}
\label{postcut}
\end{figure}

The rough procedure described in Sect.~\ref{cuts} leaves us with the light curve which is free from outliers that deviate particularly much (Fig.\,\ref{postcut}, upper panel). However, some less prominent outliers can be still identified, especially if only a small part of the data is visualized as done in the bottom panel of Fig.\,\ref{postcut} where each cloud of data points corresponds to the observations from one orbit. This example shows that working with orbit samples is an efficient way to identify outliers and {\sl this is what we would like to recommend}. The orbits usually contain enough data points to safely remove outliers on a statistical basis. Next, as we will show later, the scatter in consecutive orbits may significantly differ. This and the presence of intrinsic variability may cause problems when using samples longer than a single orbit.

What is the best tool that can identify and remove the remaining outliers in a robust way? Again, {\sl there is no single, best way to remove outliers}. There are many methods that can be used for this purpose. We have implemented the GESD algorithm\footnote{GESD stands for Generalized Extreme Studentized Deviate. The description of the algorithm can be found e.g.~at http://www.itl.nist.gov/div898/handbook/eda/section3/eda35h3.htm. It is implemented in Matlab (http://www.mathworks.com/matlabcentral/fileexchange/28501-tests-to-identify-outliers-in-data-series/content/gesd.m), but one can look for it in other packages as well. In order to support those who would like to use the GESD algorithm, a Fortran program {\tt outl-gesd.f} is provided through the BRITE Wiki page.}, which is claimed to be robust. Details of the algorithm can be found in the web page quoted in footnote \#12. One can, of course, use any other algorithm, e.g.~$\sigma$-clipping which should also work well. The GESD algorithm uses a single parameter, a level of significance $\alpha$, $\alpha >$~0. The larger $\alpha$, the more outliers are detected. How to choose this parameter optimally? The answer is not obvious and can be different for different data sets\footnote{The $\alpha$ parameter is the level of significance, but since the distribution of points for a single orbit is sometimes far from normal, $\alpha$ should be treated rather as a free parameter not a statistical parameter with its conventional meaning.}. If one works with normally distributed samples, one could use a typical value, e.g.~0.05 or 0.1. For data sets with a larger number of outliers, we would suggest using a larger $\alpha$ (the algorithm as such allows one to adopt even $\alpha>$~1). 

We have made some tests using the BAb and UBr data for $\alpha$~Cir in the Cen~I field. In general, stronger outlier removal results in smaller scatter but also in a smaller number of data points that are left. Consequently, the detection threshold in the periodogram, $D$, which depends both on the number of data points (the fewer the points, the higher is $D$) and the scatter (the smaller the scatter, the lower is $D$), can reach a minimum for a certain $\alpha$. However, the minimum in the $D(\alpha)$ relation is not always an optimal choice. There are data sets for which there is no distinct minimum in $D(\alpha)$ or the minimum occurs for large $\alpha$ corresponding to a very large number of outliers removed. We would therefore suggest {\sl not} to use the minimum of $D(\alpha)$ for choosing $\alpha$. It is better to try several values of $\alpha$ and then adopt a judiciously chosen value based on the results, the percentage of the removed data in particular. For most BRITE data sets we use a smaller $\alpha$ for BTr, BHr, and BLb data, and a slightly higher value for BAb and UBr. An example of outlier rejection for the analyzed data sample is shown in Fig.\,\ref{postcut}, where $\alpha=$ 0.3 was used.

Should the outlier rejection be applied only once or several times during the whole procedure? This is the user's decision, but we would like to recommend doing this more than once, though not too frequently. As we will see later, the decorrelation step includes many iterations. There is no need to apply outlier removal after each iteration. We usually apply a mild removal criterion (small $\alpha$ in GESD) prior to decorrelation, then again after removing the strongest one-dimensional (1D), that is, single-parameter correlations with magnitudes, and once again after finishing the two-dimensional (2D), that is, two-parameter decorrelations. The latter two removals are done with a higher $\alpha$. We proceed this way because some minor outliers often originate from strong correlations and can be corrected via decorrelations. Therefore, there is no need to remove them at an early stage. 

There are still two steps that need to be done before we start decorrelations: (i) removal of the worst orbits (those with the highest scatter) and (ii) subtraction of any strong intrinsic signal. Step (i) needs to be done because outlier removal does not remove bad orbits if data points within a bad orbit have a (nearly) normal distribution. The bad orbits frequently have mean magnitudes that differ significantly from the rest, which would affect the correlations. For the same reason, step (ii) is necessary if the intrinsic variations have large amplitudes. Note that one does not need to make a complete variability solution at this step: it is enough to remove only the strongest variability. Although we would recommend doing (i) before (ii), the reverse sequence may work as well. 

\subsection{Removal of the worst orbits}\label{worst}
\begin{figure}[!ht]
\centering
\includegraphics[width=0.8\textwidth]{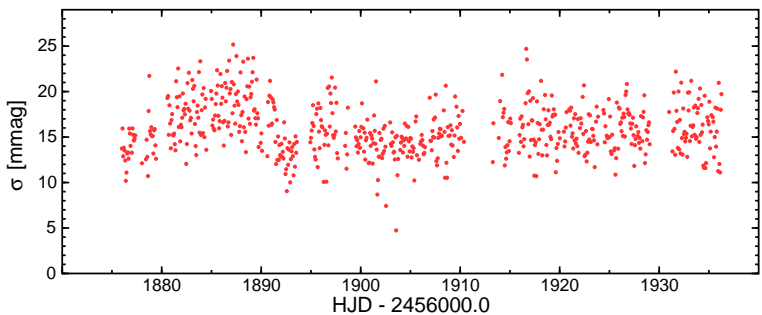}
\caption{Standard deviations, $\sigma$, for orbital samples in our example data set (BHr data of HD\,92287, setup 8).}
\label{worbits}
\end{figure}

Let us have a look at an example of step (i). Figure \ref{worbits} shows standard deviations, $\sigma$, in orbit samples for the same data as shown in Fig.\,\ref{postcut} (BHr observations of HD\,92287, setup 8). As one can see, the mean $\sigma$ amounts to about 16~mmag, but there are no orbits that have much larger scatter than the rest of data (the worst orbit has $\sigma =$ 25.2~mmag). In this case, we would not remove any orbit. Obviously, this procedure does not remove deviating points in the light curve, which are due to the intrinsic variability (e.g.~short eclipses) --- which is good. The calculated values of $\sigma$ can be used as uncertainties of the data points. Again, the removal of the worst orbits can be done more than once because decorrelations (especially if correlations are strong) considerably change $\sigma$.

\subsection{Comments on the intrinsic variability}\label{intrinsic}
Let us now move to step (ii). An important question arises at this point: should we at all remove intrinsic variability prior to decorrelations?\footnote{If the timescale of the variability is considerably shorter than the duration of the observations in a single orbit, one may consider subtracting the intrinsic variability even before outlier rejection is performed.} The answer is: it depends. If the amplitude of the intrinsic variability is much smaller than the scatter of points within a single orbit, one does not need to bother about subtracting it. What if one needs to or simply wants to do this anyway? A simple recipe can be the following:
\begin{enumerate}
\item Remove outliers in a simple way, e.g.~following the procedure presented in Sect.~\ref{cuts}. This will not remove all outliers, but those that deviate most will be removed.  In addition, a robust outlier removal with a mild rejection criterion (Sect.~\ref{rem-out})  and worst-orbits removal (Sect.~\ref{worst}) can be applied.
\item Calculate periodogram(s), identify frequencies of the intrinsic variability and fit them to the original data. If no strong intrinsic variability is detected, this step can be skipped. The fitting is easy only when dealing with periodic variability that can be well approximated by a series of sinusoids. Alternatively, for non-periodic variability, one can describe it by a different function (e.g.~by a polynomial). Subsequently, one has to work in parallel with two light curves: use residuals from the fit to identify outliers and calculate correlations, but apply the corrections to, or remove the outliers from, the original light curve. 
\end{enumerate}

\begin{figure}[!ht]
\centering
\includegraphics[width=0.8\textwidth]{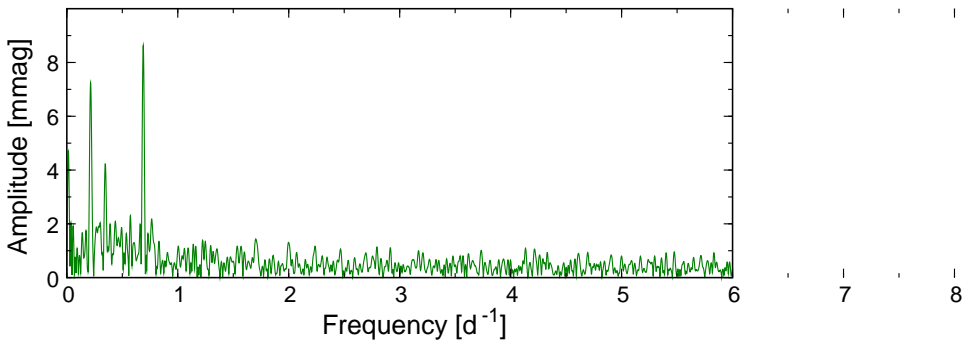}
\caption{Fourier frequency spectrum of the data used in our example (as also shown in Figs~\ref{raw-lc}\,--\,\ref{worbits}) after the first removal of outliers.}
\label{spec0}
\end{figure}

Let us come back to our example light curve shown in Fig.\,\ref{postcut}. The Fourier frequency spectrum of the data after the first outlier removal is shown in Fig.\,\ref{spec0}. One can see that there are at least three well-defined peaks with amplitudes in the range between 4 and 9~mmag\footnote{The star is classified as B3\,IV, which means that the periodicities are likely $g$ modes and this is an SPB star.}. Should we subtract them prior to the decorrelations or not? Well, since --- as we have shown in Sect.~\ref{worst} --- the average scatter per orbit is approximately 2\,--\,3 times larger than the amplitudes of the peaks in Fig.\,\ref{spec0}, this is not really necessary. However, this is the user's decision and, for the purpose of this exercise, we have fitted a three-sinusoid model to the data and will do decorrelations using residuals from this fit.

\section{Decorrelations}\label{decor}
Having removed the outliers and the worst orbits, one can perform the very important next step in this procedure, decorrelation. The raw BRITE magnitudes correlate (sometimes strongly) with temperature, position of the centroid, orbital phase and other parameters, which are calculated during the reduction and are provided with the raw data. The number of parameters increased in the subsequent Data Releases mainly as a result of the recognition that the raw data can be improved including them. In addition to the dependence on these parameters, one can expect to see instrumental effects related to the orbital phase, which is not provided with the data as a parameter and has to be calculated by the user. The orbital periods (in minutes) are given in Table 6 of Paper II. Since they can slightly evolve with time, it may be reasonable to calculate them using real data. A practical way to do this is the following. As we will show in Sect.~\ref{TSA}, if an intrinsic frequency $f_1$ is present in the data, its aliases $nf_{\rm orb} \pm f_1$ also occur, where $n$ is a natural number. The frequency $f_{\rm orb}$ and the corresponding orbital period $P_{\rm orb}= 1/f_{\rm orb}$ can therefore be derived in a simple way from $f_1$ and its nearest alias $f_1^\prime = f_{\rm orb} - f_1$, so that $f_{\rm orb} = f_1 + f_1^\prime$.

Decorrelation is, in our opinion, a critical step in the analysis of BRITE data because it can significantly improve the photometry.  The procedure we propose includes only 1D and 2D decorrelations.\footnote{In the old version of the Cookbook, there was a statement that it would be the best to make multi-dimensional decorrelation in one step. This remains valid, but is not practical. 2D correlations are in all cases much weaker than 1D correlations and it can be expected that higher-dimensional correlations would be even weaker. Moreover, decorrelations in a high-dimension (3D and higher) space pose problems due to: (i) sparsely populated parts of the multidimensional parameter space, (ii) time-consuming calculations, (iii) complicated correlations between some parameters that cannot be approximated by a simple function, (iv) inability to visualize the correlations.} Yes, {\sl there is no single best way to decorrelate BRITE data}. Since the reason for the occurrence of correlations of the raw magnitudes with CCD temperature, position of the stellar centroid and the other parameters are explained in Papers III and IV, we do not repeat them here. We will rather focus on the ways to correct for them.

As already mentioned, the 1D correlations are stronger than the 2D ones so that it makes sense to start with the 1D decorrelations. A simple recipe for a 1D decorrelation consists of three steps: (i) plot magnitudes as a function of a given parameter, (ii) fit a function which describes the dependence, and (iii) correct for the correlation by subtracting the fitted function from the data. 

Since we have at least four parameters (up to nine in DR5) to check the raw magnitudes for correlations with them, we have to decide on the sequence in which we do that. Depending on the sequence we choose, the result can be different. Based on some tests we made, the 
conclusion is as expected, namely that it is best to start with the strongest correlation and then proceed with successively weaker ones. What does `the strongest' mean? The user may decide how to evaluate the strength of the correlation. We prefer to use the parameter $R = 100(1 -  V_{\rm post}/V_{\rm pre}$), where $V_{\rm pre}$ and $V_{\rm post}$ are pre- and post-fit variances, respectively. The higher $R$, the stronger the correlation because the reduction of the variance is higher. By definition, $R$ represents the reduction (in percent)  of the variance.

The fit which we mean here is the fit of the dependency of the magnitude (residuals if intrinsic variability has been subtracted) on a given parameter. Which fitting function is the best? Experience shows that in general simple models like linear or quadratic functions do not account for the real dependencies, and are sufficient only in rare cases. Users have to decide on their own method to describe the correlations. For example, \cite{2017A&A...602A..91B} used a Bayesian technique in combination with B-splines to decide on the best fitting model. The method we chose to account for the correlations relies on the Akima interpolation \citep{1991ACMT...17..367A,1991ACMT...17..341A} between anchor points, which are calculated as averages in arbitrarily chosen intervals of a given parameter. Some examples can be found in Appendix A of \cite{2016A&A...588A..55P}. The interpolated dependence between residual magnitude and a given parameter is then subtracted from the light curve. The result forms the starting point for the next decorrelation. If the intrinsic variability is subtracted, it is fitted anew before each decorrelation step. 

\begin{figure}[!ht]
\centering
\includegraphics[width=\textwidth]{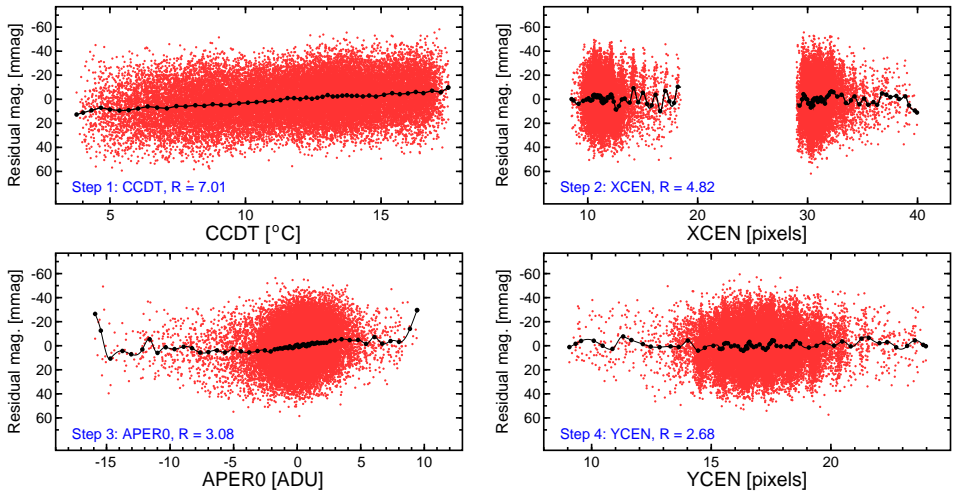}
\caption{Sequence of the first four decorrelations for the analyzed example. Each panel shows the residual magnitudes as a function of the decorrelated parameter, the anchor points (black dots) and the interpolated dependence (continuous line). The values of $R$ are also given for each step.}
\label{decor1}
\end{figure}

The sequence of decorrelations is decided on the basis of the current values of $R$, recalculated at each step of decorrelation sequence. We show the first four steps of this sequence in Fig.\,\ref{decor1}, again for the same example data (BHr data of HD\,92287, setup 8). The whole procedure stops when for all parameters $R$ falls below the user-defined value, 0.05 in our case.

As one can see, in this example the strongest correlation is for the parameter CCDT, then XCEN and so on. The full sequence of 1D decorrelations consisted in this case of 18 steps, including multiple decorrelations with some parameters (thrice each with XCEN and YCEN, twice each with CCDT, PSFC1, RTSP, APER0, and orbital phase, and once each with PSFC2 and RTSC). Writing this, we point out that multiple decorrelations with the same parameter should be applied. This is important, especially when the initial correlations are strong.

Having applied the 1D decorrelations, we continue with the 2D decorrelations. The latter describe the residual effects, but in some cases might nevertheless be strong. Applying them allows to reduce the residual standard deviation by a few percent. Again, one can describe the 2D correlation in its own way. We chose to calculate the same $R$ value as for the 1D decorrelations for all possible combinations of two parameters. This time, however, the fit of the correlation must be done in two dimensions. Instead of defining a 2D function which can be fitted, we decided to calculate a local value of the correction based on the weighted mean of the residual magnitudes from the surroundings of a selected point in the two-parameter space. The choice of weights is again a user decision; we use weights proportional to 1/($\sigma + d$), where $d$ is the normalized distance in the two-parameter space, and $\sigma$ is a user-chosen parameter, which defines the smoothness of the 2D correlation surface. The higher $\sigma$, the smoother is the surface. Its value is defined by the user based on the number of data points and their distribution in the parameter space. The same stopping criterion for the 2D decorrelations can be used, that is, $R=$ 0.05. In the analyzed example, the strongest 2D correlation was for the (XCEN, YCEN) parameters, with $R=$ 0.29. The full sequence included eight 2D decorrelations, six for (XCEN, YCEN), and one each for (XCEN, CCDT) and (YCEN, CCDT). All the other 2D correlations were negligible ($R<$ 0.05).

Finally, having finished the 2D decorrelations, one can come back and check if corrections for 2D correlations changed the 1D correlations. This sometimes happens, so that some additional 1D decorrelations may be required to meet the same criterion (e.g.~$R=$ 0.05) both for 1D and 2D decorrelations. In other words, a typical decorrelation sequence is the following: 1D --- 2D --- 1D. For the example we show here, the initial (prior to the decorrelations) residual (i.e.~after subtracting the three-mode model mentioned in Sect.~\ref{intrinsic}) standard deviation, RSD, was equal to 16.33~mmag. The whole procedure ended with RSD $=$ 14.20, that is, lower by 13.0\% with respect to the initial value. The contributions to this number from the three factors were the following: 1D decorrelations, 10.7\%, 2D decorrelations, 0.4\%, and the additional outlier removal, the remaining 1.9\%. The example we worked with does not show strong correlations. However, in other cases, decorrelations result in a 50\% or even larger drop of the RSD.

The final comment in this section is related to the following questions: Should individual setups for the same star and satellite be merged prior to decorrelations? Or should they be kept separate? In general, the more data points in the sample, the better correlations can be defined and decorrelations performed. Thus, opting for merging setups seems to be reasonable. However, this should be made with caution. First of all, different setups can be merged only if the subraster position in the CCD is the same for these setups. These positions can be read off the data headers (parameters {\tt ROIxpos} and {\tt ROIypos}). Even if the positions are the same, one has to remember that setups are reduced independently, so that different optimal apertures and other fitting parameters could be defined for them. A consequence is a magnitude offset between setups. Therefore, before merging different setups, the user has to account for these magnitude offsets. For the reason given above, splitting a setup does not seem to be a good choice, but in some very rare cases it may happen that something, which cannot be described with the correlation parameters, happened during observations and, in order to get a better result, it is reasonable to split a long setup into parts and perform decorrelations separately for each of them.

\section{Time-series analysis}\label{TSA}
Once the data are decorrelated, the corrected light curve can be subjected to a time-series analysis and scientific interpretation. This part of the work is beyond the scope of this paper, but a few comments on the alias patterns that can be expected in the frequency spectra of BRITE data might be useful for potential users. Two sampling rates determine the alias pattern of the data. The first one is related to the rate in which the consecutive images are secured. The BRITE frames are typically taken with 1-s exposures separated by 20\,--\,23~s gaps. This means that the Nyquist frequency related to this sampling is very high ($\sim$2000~d$^{-1}$) and very high frequencies can, in principle, be detected in the data. On the other hand, there are very few stars among the potential BRITE targets, in which such high frequencies are expected. The rapidly oscillating Ap star $\alpha$~Cir with its 210~d$^{-1}$ pulsation is an example \citep{2016A&A...588A..54W}. 

\begin{figure}[!ht]
\centering
\includegraphics[width=0.8\textwidth]{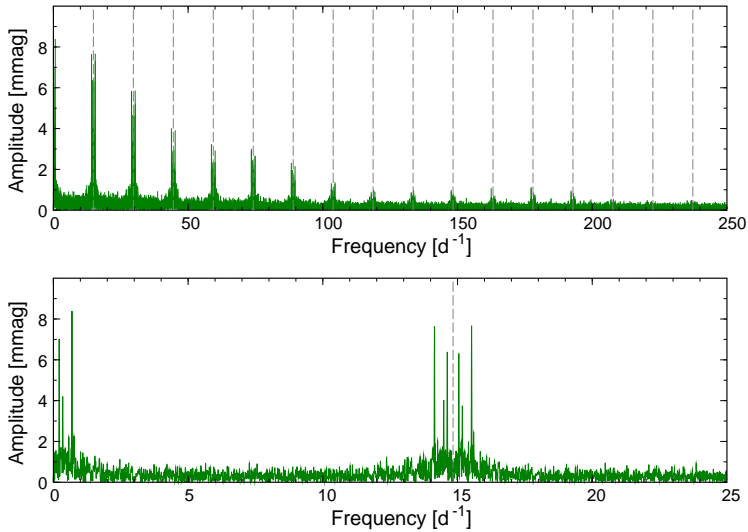}
\caption{Top: Frequency spectrum of the example data after decorrelations in the range 0--250~d$^{-1}$. Bottom: The same as above but in the range of 0--25~d$^{-1}$. Vertical dashed lines stand for the orbital frequency $f_{\rm orb}$ and its multiples.}
\label{nyquist}
\end{figure}

The other sampling rate is related to the orbital periods of the BRITEs, which amount to about 100 minutes (the corresponding orbital frequency $f_{\rm orb}\approx$~14.4~d$^{-1}$). The resulting Nyquist frequency $f_{\rm N,orb}=f_{\rm orb}/\mbox{2}\approx$  7.2~d$^{-1}$. Let us have a look at the frequency spectrum of the data we used as an example (Fig.\,\ref{nyquist}), but this time in a  wider range of frequencies, much higher than $f_{\rm N,orb}$. 

As already explained in Sect.~\ref{decor}, the aliases of an intrinsic frequency $f_i$ occur at frequencies $nf_{\rm orb} \pm f_i$, where $n$ is an integer number. The aliases can be seen even at very high frequencies. The envelope of the peaks is simply described by the absolute value of the sinc ($\sin x/x$) function related to the width of the observations made during a single orbit (about 12.5 minutes). It is obvious that strong aliases may pose a problem for the unambiguous identification of the intrinsic frequencies if the latter are close to $f_{\rm N,orb}$. This problem can be solved by combining data from more than one satellite, if available, even if this requires a combination of blue- and red-filter data. Such a procedure allows not only to reduce the aliasing, but also to lower the detection threshold and therefore find low-amplitude periodic variability. An excellent example of the power of combining BRITE data from different satellites and of BRITE and ground-based data has been presented by \cite{2017MNRAS.464.2249H} for $\nu$~Eri\footnote{\cite{2017MNRAS.464.2249H} presented a more sophisticated method in which prewhitening was performed separately for blue- and red-filter data, but the search for significant peaks was done using combined residuals from both fits.}. Therefore, we strongly recommend combining BRITE data, especially when  low-amplitude periodic terms are searched for.

\section{Final remarks}
The present document indicates the problems we have encountered when working with BRITE data and suggests some solutions. It also includes some remarks and warnings. Similarly to the previous version of the Cookbook, we do not provide a complete script with all programs that are needed to get the final result. In fact, we hesitated whether to do this or not. We decided against. It is better to go through the whole procedure step by step and understand what is going on instead of using somebody else's program as a black box. Nevertheless, some programs which can help to build the user's own script will be made available through the BRITE Wiki page. Finally, the BRITE team offers its expertise and help in working with BRITE data. Then, good luck and have fun working with them!

\acknowledgements{We would like to thank very much all members of the BRITE Team (especially the members of the PHOTT and QCT teams) who contributed to this document with all the ideas and insights expressed in many discussions: Bram Buysschaert, Gerald Handler, Rainer Kuschnig, Tony Moffat, Melissa Munoz, Bert Pablo, Adam Popowicz, Tahina Ramiaramanantsoa, Jason Rowe, S{\l}awek Ruci\'nski, Radek Smolec, Werner Weiss, Gemma Whittaker, and Ela Zoc{\l}o\'nska. The study is based on data collected by the BRITE Constellation satellite mission, designed, built, launched, operated and supported by the Austrian Research Promotion Agency (FFG), the University of Vienna, the Technical University of Graz, the Canadian Space Agency (CSA), the University of Toronto Institute for Aerospace Studies (UTIAS), the Foundation for Polish Science \& Technology (FNiTP MNiSW), and National Science Centre (NCN). The operation of the Polish BRITE satellites is secured by a SPUB grant of the Polish Ministry of Science and Higher Education (MNiSW). This work was supported by the NCN grant no.~2016/21/B/ST9/01126.}


\begin{thebibliography}{9}
\providecommand{\natexlab}[1]{#1}
\providecommand{\url}[1]{\texttt{#1}}
\providecommand{\urlprefix}{URL }
\providecommand{\eprint}[2][]{\url{#2}}

\bibitem[{{Akima}(1991{\natexlab{a}})}]{1991ACMT...17..367A}
{Akima}, H., \emph{Algorithm 697: Univariate interpolation that has the
  accuracy of a third-degree polynomial}, \emph{ACM Transactions on
  Mathematical Software} \textbf{17}, 3, 367 (1991{\natexlab{a}})

\bibitem[{{Akima}(1991{\natexlab{b}})}]{1991ACMT...17..341A}
{Akima}, H., \emph{A method for univariate interpolation that has the accuracy
  of a third-degree polynomial}, \emph{ACM Transactions on Mathematical
  Software} \textbf{17}, 3, 341 (1991{\natexlab{b}})

\bibitem[{{Buysschaert} et~al.(2017)}]{2017A&A...602A..91B}
{Buysschaert}, B., et~al., \emph{{Studying the photometric and spectroscopic
  variability of the magnetic hot supergiant {$\zeta$} Orionis Aa}},
  \emph{\aap} \textbf{602}, A91 (2017)

\bibitem[{{Handler} et~al.(2017)}]{2017MNRAS.464.2249H}
{Handler}, G., et~al., \emph{{Combining BRITE and ground-based photometry for
  the {$\beta$} Cephei star {$\nu$} Eridani: impact on photometric pulsation
  mode identification and detection of several g modes}}, \emph{\mnras}
  \textbf{464}, 2249 (2017)

\bibitem[{{Pablo} et~al.(2016)}]{2016PASP..128l5001P}
{Pablo}, H., et~al., \emph{{The BRITE Constellation nanosatellite mission:
  testing, commissioning, and operations}}, \emph{\pasp} \textbf{128}, 12,
  125001 (2016), {(Paper II)}

\bibitem[{{Pigulski} et~al.(2016)}]{2016A&A...588A..55P}
{Pigulski}, A., et~al., \emph{{Massive pulsating stars observed by
  BRITE-Constellation. I. The triple system {$\beta$} Centauri (Agena)}},
  \emph{\aap} \textbf{588}, A55 (2016)

\bibitem[{{Popowicz} et~al.(2017)}]{2017A&A...605A..26P}
{Popowicz}, A., et~al., \emph{{BRITE Constellation: data processing and
  photometry}}, \emph{\aap} \textbf{605}, A26 (2017), {(Paper III)}

\bibitem[{{Weiss} et~al.(2014)}]{2014PASP..126..573W}
{Weiss}, W.~W., et~al., \emph{{BRITE-Constellation: nanosatellites for
  precision photometry of bright stars}}, \emph{\pasp} \textbf{126}, 573
  (2014), {(Paper I)}

\bibitem[{{Weiss} et~al.(2016)}]{2016A&A...588A..54W}
{Weiss}, W.~W., et~al., \emph{{The roAp star {$\alpha$} Circinus as seen by
  BRITE-Constellation}}, \emph{\aap} \textbf{588}, A54 (2016)

\end{thebibliography}
\end{document}